# Magnetic exchange induced Weyl state in a semimetal EuCd$_2$Sb$_2$

Hao Su[1,2†], Benchao Gong[3†], Wujun Shi[1†], Haifeng Yang[1], Hongyuan Wang[1,2], Wei Xia[1,2], Zhenhai Yu[1], Peng-Jie Guo[3], Jinhua Wang[4], Linchao Ding[4], Liangcai Xu[4], Xiaokang Li[4], Xia Wang[5], Zhiqiang Zou[5], Na Yu[5], Zengwei Zhu[4], Yulin Chen[1,6], Zhongkai Liu[1\*], Kai Liu[3\*], Gang Li[1\*], Yanfeng Guo[1,2\*]

[1]*School of Physical Science and Technology, ShanghaiTech University and CAS-Shanghai Science Research Center, Shanghai 201210, China*

[2]*University of Chinese Academy of Sciences, Beijing 100049, China*

[3]*Department of Physics and Beijing Key Laboratory of Opto-electronic Functional Materials & Micro-nano Devices, Renmin University of China, Beijing 100190, China*

[4]*Wuhan National High Magnetic Field Center and School of Physics, Huazhong University of Science and Technology, Wuhan 430074, China*

[5]*Analytical Instrumentation Center, School of Physical Science and Technology, ShanghaiTech University, Shanghai 201210, China*

[6]*Department of Physics, University of Oxford, Oxford, OX1 3PU, United Kingdom*

[†]The authors contributed equally to this work.

[\*]E-mails:

liuzhk@shanghaitech.edu.cn;

kliu@ruc.edu.cn;

ligang@shanghaitech.edu.cn;

guoyf@shanghaitech.edu.cn.

**Magnetic Weyl semimetals (WSMs) bearing long-time pursuing are still very rare. We herein identified magnetic exchange induced Weyl state in EuCd$_2$Sb$_2$, a semimetal in type IV magnetic space group, via performing high magnetic field**

**(B) magneto-transport measurements and *ab initio* calculations. For the A-type antiferromagnetic (AFM) structure of $EuCd_2Sb_2$, external B larger than 3.2 T can align Eu spins to be fully polarized along the *c*-axis and consequently drive the system into a ferromagnetic (FM) state. Measurements up to B ~ 55 T revealed a striking Shubnikov-de Hass oscillation imposed by a nontrivial Berry phase. We unveiled a phase transition from a small-gap AFM topological insulator into a FM WSM in which Weyl points emerged along the Γ-Z path. Fermi arcs on (100) and (010) surfaces are also revealed. The results pave a way towards realization of various topological states in a single material through magnetic exchange manipulation.**

A recent research frontier in condensed matter physics is the realization of various fermionic particles and related phenomena in solids, naturally bridging with those predicated in high-energy physics[1]. Three varieties of fermions in solids, i.e. Dirac[2-4], Weyl[5-7], and Majorana[8-16], have been chronologically realized in topological semimetals (TSMs) and superconductors and have captured immense interest because of the associated exotic quantum features[17-23]. Three-dimensional (3D) TSMs hosting Dirac and Weyl fermions are characterized by nontrivial bulk band crossing which forms discrete nodal points protected by certain symmetry against gap formation. The low-energy excitation around the nodal points is viewed as quasiparticles or emergent relativistic fermions that can be satisfyingly described by the massless relativistic Dirac/Weyl equation behaving as an electronic band linear dispersion along all three momentum directions[2-7, 24, 25]. The nodal point in 3D Dirac semimetal (DSM) is fourfold degenerate formed by the crossing of two doubly degenerate bands[2-4, 24, 25]. Once the spin degeneracy is lifted by either breaking the time-reversal symmetry (*T*) or spatial inversion symmetry (*P*), the Dirac point (DP) characterized by chiral symmetry will split into a pair of Weyl points (WPs) which behave as the monopoles of Berry curvature[5-7]. The WPs emerged in pairs are well separated in momentum space under the conservation of the opposite chirality of the Berry curvature, which consequently are topologically stable and will not be annihilated unless they could be moved to the same $\kappa$-point in the Brillouin zone (BZ). The existence of WPs can produce extraordinary intriguing properties, such as chiral anomaly effect, gravitational effect, strong intrinsic anomalous and spin Hall effect, large magnetoresistance (MR), etc.[19-23].

The *P*-breaking Weyl fermions, which is realized entirely by crystal structure, have been established in a number of Weyl semimetals (WSMs) such as TaAs family[6, 7, 26], (W/Mo)Te$_2$ [27, 28], and photonic crystals[29], etc. As a contrast, *T*-breaking Weyl fermions are still very rare despite numerous materials have been theoretically predicated[30-33]. The experimental detection of the WPs in these materials has encountered difficulties mainly in that the WPs are far away from the Fermi level ($E_F$) or charge carrier density of trivial Fermi surfaces (FSs) is very large. Thus more ideal WSMs are strongly desired for the Weyl physics. To find a *T*-breaking WSM, one can either apply a finite external magnetic field (B), for example, in GdPtBi[34], or use the spontaneous magnetism to break *T*, for example, in YbMnBi$_2$[32]. However, these proposed materials are still waiting for solid evidences for the identification of the Weyl states. We herein report a very different way to create *T*-breaking Weyl fermions, that is, in EuCd$_2$Sb$_2$ with the type IV magnetic space group, via switching the antiferromagnetic (AFM) structure into a ferromagnetic (FM) one by applying external magnetic field to manipulate the spin configuration.

Details of EuCd$_2$Sb$_2$ crystal growth, basic characterizations on the crystal structure, phase quality, magnetization, and resistivity are presented in Figures S1-S2 of the supplementary information (SI). The results are harmonious with those in earlier reports[35,36]. For the A-type AFM structure hosting in-plane magnetic moments which form FM layers stacking antiferromagnetically along the *c*-axis, seen by the sketch views in Figures S1(b)-S1(d), a small B (~3.2 T) is sufficient to completely align the spins along the *c*-axis, i.e. switching the system into a FM state. The isothermal magnetization (M(B)) depicted in the inset of Figure S2(a) reveals a saturation moment close to 7 $\mu_B$ f.u.$^{-1}$ at 2 K and B> 3.2 T. Unlike the two-dimensional (2D) DSM EuMnBi$_2$[37], we did not observe any spin-flop transitions in EuCd$_2$Sb$_2$.

The temperature (T) dependent longitudinal resistivity $\rho_{xx}$ measured at various B presented in Figure S2(b) demonstrate not only a semi-metallic behavior but also the influences from the AFM ordering and the localized Eu$^{2+}$. Since the easy alignment of Eu spins along the *c*-axis in the AFM structure by B, it naturally reminds us of the magneto-transport measurements on EuCd$_2$Sb$_2$ with B//*c* to experimentally map out the FS in the FM structure. The B dependence of $\rho_{xx}$ measured up to 55 T in pulsed magnetic fields applied along the *c*-axis with current (I) in the *ab*-plane (B//*c*⊥I) and at T = 2 - 30 K is depicted in Figure 1(a). The Shubnikov-de Hass oscillations (SdH) are clearly visible at large B by subtracting a cubic polynomial background from the experimental data. The extracted relative amplitudes of the SdH oscillations plotted against the reciprocal magnetic field 1/B are presented in Figure 1(b). As can be inferred from this figure, the SdH oscillations which are striking at B > 14 T remain discernible at temperatures up to at least 30 K but their amplitudes systematically decrease with increasing the temperature. The fast Fourier transform (FFT) of the SdH oscillations, shown in Figure1(c), discloses two features at the two fundamental frequencies, 30 T and 117 T. Each FFT peak at the fundamental or higher harmonics can calibrate the external FS area which is perpendicular to the magnetic field[38], so the two peaks likely could be assigned to two-hole FSs (marked as $\alpha_1$ and $\alpha_2$) since the transverse Hall resistivity $\rho_{xy}$ as a function of temperature and B parallel to the *c*-axis (the inset in Figure 1(a)) has a positive slope. Assuming circular FS cross-section of the hole bands within the *ab*-plane, the external cross-sectional area $A_F$ of the FS could be evaluated by using the Onsager relation $F = (\hbar/2\pi e)A$, which gives $A_{F1} = 1.11(4)$ nm$^{-2}$ and $A_{F2} = 0.028(5)$ nm$^{-2}$, corresponding to the frequencies of 117 T and 30 T, respectively. However, theoretical calculations exposed two hole and two electron bands in the FM structure of EuCd$_2$Sb$_2$ rather than only two hole bands crossing $E_F$, seen in Figure S6 by the sketch maps in which two hole-type pocket around Γ point and two electron-type pockets around Z point can be found, likely inconsistent with the experiment. The inconsistence might be interpreted as that the oscillations of rather small FSs of the two electron bands are too weak to be detected by magneto-transport measurements. The effective mass $m^*$ of quasi-particles on the Fermi pocket could be obtained by fitting the temperature dependence of the FFT amplitude of the SdH oscillations by the temperature damping factor of the Lifshitz-Kosevich (L-K) equation[39], that is, $\Delta\rho_{xx} \propto X/\sinh(X)$, where $X = 14.69 m^* T/B'$ and $B'$ is the average field. The fitted results shown in Figure 1(d) by the main panel give cyclotron mass $m^*$ of $0.237 m_0$ for the 117 T hole pocket and $0.355 m_0$ for the 30 T hole pocket, where $m_0$ denotes the mass of a free electron. The fitted results to $\rho_{zz}$ in Figure S4(d) also give the harmonic values for the two hole-pockets, suggesting the reliability of the analysis. The two values are larger than that of the gapless DSM $Cd_3As_2$ but are comparable with that of $SrMnBi_2$[40,41]. The analysis also yields the Fermi vectors $k_F$ of 0.596 nm$^{-1}$ and 0.302 nm$^{-1}$ for the 117 T and 30 T hole pockets, respectively, associated with very large fermion velocities $v_F = 2.9 \times 10^5$ m s$^{-1}$ and $9.8 \times 10^4$ m s$^{-1}$, respectively, which could consequently be estimated through $v_F = \hbar k_F / m^*$ in the case of an ideal linear dispersion.

To provide more evidences for demonstrating the nontrivial topological state in in $EuCd_2Sb_2$ at high B, the Berry phase $\varphi_B$ accumulated along cyclotron orbits was examined. Generally speaking, the pseudo-spin rotation under a magnetic field in a Dirac/Weyl system will produce a non-trivial $\varphi_B$ which could be accessed from the Landau level (LL) index fan diagram or a direct fit of SdH oscillations by using the L-K formula[42-44]. The phase shift is generally a sum $\varphi_i = -\gamma+\delta$, where $\gamma$ is the phase factor expressing as $1/2-\varphi_B/2\pi$ and $\delta$ represents the dimension-dependent correction to the phase shift[39]. In a 2D case, this parameter amounts zero, while in a 3D case $\delta$ is equal to $\pm 1/8$ where the sign depends on type of charge carriers and kind of cross-section extremum[45,46]. The inset in Figure 1(d) shows the plot of N as a function of 1/B for the 117 T hole pocket. Here the $\Delta\rho_{xx}$ valley positions (closed circles) in 1/B were assigned to be integer indices and the $\Delta\rho_{xx}$ peak positions (open circles) were assigned to be half-integer indices. All the points almost fall on a straight line, which allows a linear fitting that gives an intercept 0.41(1), demonstrating a nonzero $\varphi_B$. Importantly, since the lowest observed Landau index is 3 T at 45 T, the uncertainty associated with the fitted intercept is very small, indicating that the determination of $\varphi_B$ is reliable. The B dependence of $\rho_{zz}$ up to 60 T measured with the B//*c*//I configuration at T = 2 - 30 K are presented in Figure S4 along with analysis results. The basic conclusions suggesting nontrivial Berry phase for both frequencies at 30 T and 117 T are similar as those given above, further guarantee the reliability of our analysis.

The electronic structures of $EuCd_2Sb_2$ from first-principles calculations at AFM and FM states are shown in Figure 2. Details of the calculations can be found in the SI. Below the Neel temperature $T_N \sim 7.5$ K, $EuCd_2Sb_2$ is AFM-ordered with Eu spins lying in the *ab* plane[6]. Due to the broken $C_3$ symmetry along *c*-axis, the Dirac point (DP) near the $E_F$ along Γ-Z path of BZ shown in Figure 2(a) is no longer protected by the combined inversion and non-symmorphic time-reversal symmetries [36, 47, 48], but rather opens tiny band gaps as is seen from Figure 2(c), making $EuCd_2Sb_2$ a topological insulator with a momentum-dependent chemical potential. The topology is further confirmed from the evaluation of the $Z_2$ invariant protected by the joint operation of *P* and non-symmorphic *T* symmetries, where the latter alone is broken by the AFM. Since $EuCd_2Sb_2$ owns inversion symmetry, the topological invariant $Z_2$ can be calculated by the parity product of all occupied states at the time-reversal invariant momenta (Table SI), which yields 1. Once B > $B_c$, all spins of Eu ions become polarized along the *c*-axis and consequently form an FM long range order, which breaks the *PT* symmetry of the AFM phase resulting in a completely different topology of $EuCd_2Sb_2$. The FM electronic structure is shown in Figure 2(d), in which WPs near $E_F$ can be clearly observed. We note that all the WPs in this system are along the Γ-Z path. Focusing on an energy window of -0.1 - 0.1 eV around $E_F$, we found five pairs of WPs with their specific positions and energies in the BZ summarized in Table S2. Moreover, the carrier concentrations ($n_H = 3.3 \times 10^{19} cm^{-3}$) calculated from the FS volumes of the two hole pockets are in good accordance with the value ($n_H = 3.13 \times 10^{19} cm^{-3}$) calculated by the Hall resistivity shown by the inset in Figure 1(a).

The nontrivial band structure and susceptible magnetism in EuCd$_2$Sb$_2$ are intimately related to the transport properties. For Dirac/Weyl semimetal, when B//I, the negative MR can appear due to a population imbalance between Weyl fermions of different charities, which can produce a net electric current[21]. At relatively low magnetic field (< 2 T), the magnetic structure is A-type AFM with tiny gaps which is rather close to be a DSM (Figures 2(b) and 2(c)). Thus, as seen in Figure 3(b), the B dependent of the negative MR can be nicely fitted with the chiral magnetic effect within the range of -0.5 T ~ 0.5 T, given by $\sigma(B) = (1 + C_w B^2)(\sigma_0 + a\sqrt{B}) + \sigma_N$, where $C_w B^2$ represents the contribution of the chiral current with a positive value, $\sigma_0$ is the zero-field conductivity, $a\sqrt{B}$ is the weak antilocalization contribution with a negative value, and $\sigma_N$ denotes the conventional nonlinear band contribution around the $E_F$[49]. The failed fit to the data within 0.5 T – 2 T is caused by the more prominently changed magnetic structure in which the Eu spins are progressively polarized along the $c$-axis with the increase of B. When B > 2 T, the polarization gradually reach its saturation, the electronic structure thereby reconstructs with more bands crossing the $E_F$ as is seen in Figure 2(d). The reduced scattering by parallel spins along $c$-axis would further diminish the resistance, resulting in a down-turn of the negative MR for both B//I and B⊥I as is seen in Figure 3(a). However, the critical points of the MR for B//I and B⊥I starting to rise for are different, 6.7 T and 4.5 T, respectively, which could be attributed to the effects from chiral anomaly of WPs. The combination of chiral anomaly and ferromagnetism in EuCd$_2$Sb$_2$ can thus cause more prominent negative MR than that in nonmagnetic Weyl semimetal TaAs[12], while the weak inter-layer magnetic coupling in EuCd$_2$Sb$_2$ provides an additional freedom to tune the transport properties.

The characteristic feature of the Weyl systems lies in their existence of Fermi arc connecting a pair of nodes with opposite chirality. The WPs of EuCd$_2$Sb$_2$ around the

$E_F$ embed in the other bulk bands generally hindering the observation of their arc as they would ultimately hybridize with the bulk bands. Fortunately, some pairs of the WPs in EuCd$_2$Sb$_2$ clearly demonstrate visible arcs spreading over a large territory of the surface BZ which was not often seen in other WSMs. The calculated states on the (100) and (010) surfaces are illustrated in Figures 4(a-d). These two surfaces are different, as EuCd$_2$Sb$_2$ is only 3-fold rotationally invariant about *c*-axis. As clearly seen from the surface states comparison shown in Figures 4 (a, c) and (b, d), although the bulk states (red shaded area) demonstrate strong similarity the states originating from the surfaces (red lines) behave clearly different. Beside some clear trivial surface bands forming closed circles around the vertical BZ boundary, other surface states stably reside on both surface BZs despite of the different surface potentials, indicating their topological stability. Although they hybridize with the bulk bands around Γ making the identification of clear Fermi arc difficult, nevertheless one can still attribute them to different pairs of WPs. The observation of Fermi arcs further evidences the WPs appeared in the FM structure. Generally, one pair of Dirac points will split into two pairs of WPs when *T* is broken, the emergence of the WPs in EuCd$_2$Sb$_2$ could be interpreted by the fact that in the FM state the band structure reconstructs with emerging topology (Figure 2(d)), which is not just due to the lift of band degeneracy.

In summary, by performing high magnetic field magneto-transport measurements and *ab* initio calculations, we demonstrate a rare example showing AFM topological insulator to FM *T*-breaking WSM transition in a single material EuCd$_2$Sb$_2$ belonging to the type IV magnetic space group. The appearance of *T*-breaking WPs induced by such a magnetic exchange change has never been observed in other materials. It provides an ideal material not only for investigating the relation between magnetic order and band topology, but also demonstrates a system in realization of various topological states via manipulating the magnetic exchange. Interestingly, the FM WSM provides the opportunities to study other exotic physics, such as the anomalous Nernst and thermal Hall effects, etc.[50-53]. The further investigations on this type of materials will definitely open new horizons for us in understanding the band topology theory as well as in using them in novel devices.

*Note added*: During preparation of this paper, we became aware of two related work appeared on arXiv. One is a theoretical predication of single pair of WPs in $EuCd_2As_2$ but lacks of experiments[56]. The other one reported the theoretical and experimental studies on our $EuCd_2As_2$ crystals, which claimed the discovery of an ideal Weyl state induced by magnetic exchange[57]. The information about the Fermi arcs that are crucial for a WSM was not found. Moreover, our investigation on the different system $EuCd_2Sb_2$ was totally independent and we have no private communications about our work during the whole process.

**Methods**

**Single-crystal growth, compositions and single crystal x-ray diffraction chracterizations.** The $EuCd_2Sb_2$ single crystals were grown by using tin as the flux. Starting materials including europium (99.9%), cadmium (99.999%), antimony (99.999%) and tin (99.999%) were mixed in a molar ratio of Eu : Cd : Sb : Sn = 1: 2: 2: 4. The mixture was sealed in a quartz tube, heated up to 1000 ℃ in a furnace and kept at the temperature for 30 hours, then cooled down slowly to 450 ℃ at rate of 2 K/h. The assembly was finally taken out of the furnace at 450 ℃ and was put into a centrifuge immediately to remove the excess flux. Compositions of the crystals were examined by using energy-dispersive X-ray (EDX) spectroscopy. The phase and quality examinations of $EuCd_2Sb_2$ were performed on a single-crystal x-ray diffractometer equipped with a Mo K$\alpha$ radioactive source ($\lambda$ = 0.71073 Å).

**Magnetization and magneto-transport measurements.**

Magnetic properties of $EuCd_2Sb_2$ were characterized on a commercial magnetic property measurement system (MPMS-3) from Quantum Design Company. The temperature dependent magnetization (M(T)) along the out-of-plane (B//*c*) direction was measured in the zero-field cooling (ZFC) and field-cooling (FC) mode between 1.8-300 K. The temperature dependence of longitudinal resistivity $\rho_{xx}$ measured by a four-probe configuration up to 9 T in a commercial physical property measurement system (PPMS) apparatus at various B. High-field electrical transport measurements were carried out on a pulsed magnet of 50 ms in Wuhan National High Magnetic Field Center in the field range of -55 T to 55T.

**ARPES measurements and the Hubbard U determination.**

High-resolution angle-resolved photoemission spectroscopy (ARPES) was used to directly measure the band structure of the paramagnetic $EuCd_2Sb_2$. ARPES data was mainly acquired at BL4 and BL10 of ALS, USA, at 20 K. The overall energy and angle resolutions were 15 meV and 0.2°, respectively. Fresh $EuCd_2Sb_2$ surfaces for ARPES measurements were obtained by *in-situ* cleaving the crystals at low temperatures (20 K).

**The *ab initio* calculations**

The first-principles electronic structure calculations on $EuCd_2Sb_2$ were carried out by using the projector augmented wave (PAW) method[58] as implemented in the VASP package[59]. The generalized gradient approximation (GGA) of Perdew-Burke-Ernzerhof [60] was employed for the exchange-correlation functional. The kinetic energy cutoff of the plane-wave basis was set to be 340 eV. The experimental lattice constants and atomic positions of $EuCd_2Sb_2$ were adopted[36]. A $16 \times 16 \times 10$ *k*-point mesh for the BZ sampling of the primitive cell and the Gaussian smearing technique with a width of 0.05 eV for the FS broadening were utilized. The spin-orbit coupling (SOC) effect was taken into consideration. The correlation effect among Eu 4*f* electrons was incorporated by using the GGA+U formalism of Dudarev *et al.*[61] with an effective Hubbard interaction U. According to a previous study[36], the energies of Eu 4*f* bands relative to the $E_F$ are sensitive to the Hubbard interaction. We herein chose the Hubbard U of 4.5 eV as is discussed in the SI. The maximally localized Wannier functions method[62, 63] was used to calculate the

FSs, from whose volumes the carrier concentrations were obtained.

**Data availability.** The data that support the plots within this paper and other findings of this study are available from the corresponding authors upon reasonable request.

**References**

[1] P. B. Pal, American Journal of Physics **79**, 485 (2011).

[2] Z. K. Liu, J. Jiang, B. Zhou, Z. J. Wang, Y. Zhang, H. M. Weng, D. Prabhakaran, S. K. Mo, H. Peng, P. Dudin, T. Kim, M. Hoesch, Z. Fang, X. Dai, Z. X. Shen, D. L Feng, Z. Hussain, and Y. L. Chen, Nat. Mater. **13**, 677 (2014).

[3] S.-Y. Xu, C. Liu, S. K. Kushwaha, R. Sankar, J. W. Krizan, I. Belopolski, M. Neupane, G. Bian, N. Alidoust, T.-R. Chang, H.-T. Jeng, C.-Y. Huang, W.-F. Tsai, H. Lin, P. P. Shibayev, F.-C. Chou, R. J. Cava, and M. Z. Hasan, Science **347**, 294(2015).

[4] Z. K. Liu, B. Zhou, Y. Zhang, A. J. Wang, H. M. Weng, D. Prabhakaran, S.-M. Mo, Z. X. Shen, Z. Fang, X. Dai, Z. Hussain, and Y. L. Chen, Science **343**, 864 (2014).

[5] H. M. Weng, C. Fang, Z. Fang, B. A. Bernevig, and X. Dai, Phys. Rev. X **5**, 011029 (2015).

[6] S.-Y. Xu, I. Belopolski, N. Alidoust, M. Neupane, G. Bian, C.-L. Zhang, R. Sankar, G. Chang, Z.-J. Yuan, C.-C. Lee, S.-M. Huang, H. Zheng, J. Ma, D. Sanchez, B.-K. Wang, A. Bansil, F.-C. Chou, P. P. Shibayev, H. Lin, S. Jia, and M. Z. Hasan, Science **349**, 613 (2015).

[7] B. Q. Lv, H. M. Weng, B. B. Fu, X. P. Wang, H. Miao, J. Ma, P. Richard, X. C. Huang, L. X. Zhao, G. F. Chen, Z. Fang, X. Dai, T. Qian, and H. Ding, Phys. Rev. X **5**, 031013 (2015).

[8] L. Fu and C. L. Kane, Phys. Rev. Lett. **100**, 096407 (2008).

[9] J. P. Xu, M. X. Wang, Z. L. Liu, J. F. Ge, X. Yang, C. Liu, Z. A. Xu, D. Guan, C. L. Gao, D.-D. Guan, D. Qian, Y. Zhou, L. Fu, S.-C. Li, F.-C. Zhang, and J.-F. Jia, Phys. Rev. Lett. **116**, 257003 (2016).

[10] K. T. Law, P. A. Lee, T. K. Ng, Phys. Rev. Lett. **103**, 237001 (2009).

[11] S. M. Albrecht, A. P. Higginbotham, M. Madsen, F. Kuemmeth, T. S. Jespersen, J.

Nygård, P. Krogstrup, C. M. Marcus, Nature **531**, 206 (2016).

[12] H. Zhang, C.-X. Liu, S. Gazibegovic, D. Xu, J. A. Logan, G. Wang, N.van Loo, J. D. S. Bommer, M. W. A. de Moor, D. Car, R. L. M. Op Het Veld, P. J. van Veldhoven, S. Koelling, M. A.Verheijen, M. Pendharkar, D. J. Pennachio, B.Shojaei, J. S. Lee, C. J. Palmstrøm, E. P. A. M. Bakkers, S. D. Sarma, L. P. Kouwenhoven, Nature **556**,74 (2018).

[13] S. Jeon, Y. Xie, J. Li, Z. Wang, B. A. Bernevig, A. Yazdani, Science **358**, 772 (2017).

[14] P. Zhang, K. Yaji, T. Hashimoto, Y. Ota, T. Kondo, K. Okazaki, Z. Wang, J. Wen, G. D. Gu, H. Ding, S. Shin, Science **360**, 182 (2018).

[15] D. F. Wang, L. Y. Kong, P. Fan, H. Chen, S. Y. Zhu, W. Y. Liu, L. Cao, Y. J. Sun, S. X. Du, J. Schneeloch, R. D. Zhong, G. D. Gu, L. Fu, H. Ding, H. J. Gao, Science **362**, 333 (2018).

[16] Q. Liu, C. Chen, T. Zhang, R. Peng, Y.-J. Yan, C.-H.-P. Wen, X. Lou, Y.-L. Huang, J.-P. Tian, X.-L. Dong, G.-W. Wang, W.-C. Bao, Q.-H. Wang, Z.-P. Yin, Z.-X. Zhao, and D.-L. Feng, Phys. Rev X **8**, 041056 (2018).

[17] T. Liang, Q. Gibson, M. N. Ali, R. Cava, N. Ong, Nat. Mater. **14**, 280 (2015).

[18] M. N. Ali, J. Xiong, S. Flynn, J. Tao, Q. D. Gibson, L. M. Schoop, T. Liang, N. Haldolaarachchige, M. Hirschberger, N. P. Ong, and R. J. Cava, Nature **514**, 205 (2014).

[19] K.-Y. Yang, Y.-M. Lu, Y. Ran, Phys. Rev. B **84**, 075129 (2011).

[20] S. Zhong, J. Orenstein, J. E. Moore, Phys. Rev. Lett. **115**, 117403 (2015).

[21] H. B. Nielsen, M. Ninomiya, Phys. Rev. Lett. **130**, 389 (1983).

[22] J. Xiong, S. K. Kushwaha, T. Liang, J. W. Krizan, M. Hirschberger, W. Wang, R. J. Cava, N. P. Ong, Science **350**, 413 (2015).

[23] X. C. Huang, L. X. Zhao, Y. J. Long, P. P. Wang, D. Chen, Z. H. Yang, H. Liang, M. Q. Xue, H. M.Weng, Z. Fang, X. Dai, G. F. Chen, Phys. Rev. X **5**, 031023 (2015).

[24] Z. J. Wang, Y. Sun, X. Q. Chen, C. Franchini, G. Xu, H. M. Weng, X. Dai, Z. Fang, Phys. Rev. B **85**, 195320 (2012).

[25] Z. J. Wang, H. M. Weng, Q. S. Wu, X. Dai, Z. Fang, Phys. Rev. B **88**, 125427

(2013).

[26] L. X. Yang, Z. K. Liu, Y. Sun, H. Peng, H. F. Yang, T. Zhang, B. Zhou, Y. Zhang, Y. F. Guo, M. Rahn, D. Prabhakaran, Z. Hussain, S.-K. Mo, C. Felser, B. Yan, and Y. L. Chen, Nat. Phys. **11**, 728 (2015).

[27] Y. Wu, D. Mou, N. H. Jo, K. Sun, L. Huang, S. L. Bud'ko, P. C. Canfield, and A. Kaminski, Phys. Rev. B **94**, 121113 (2016).

[28] K. Deng, G. Wang, P. Deng, K. Zhang, D. Ding, E. Wang, M. Yan, H. Huang, H. Zhang, Z. Xu, J. Denliger, A. Fedorov, H. Yang, W. Duan, H. Yao, Y. Wu, S. Fan, H. Zhang, X. Chen, and S. Zhou, Nat. Phys. **12**, 1105 (2016).

[29] L. Lu, Z. Y. Wang, D. X. Ye, L. X. Ran, L. Fu, J. D. Joannopoulos, M. Solijačić, Science **349**, 622 (2015).

[30] X. Wan, A. M. Turner, A. Vishwanath, and S. Y. Savrasov, Phys. Rev. B **83**, 205101 (2011).

[31] G. Xu, H. Weng, Z. Wang, X. Dai, and Z. Fang, Phys. Rev. Lett. **107**, 186806 (2011).

[32] S. Borisenko, D. Evtushinsky, Q. Gibson, A. Yaresko, T. Kim, A. N. Ali, B. Buechner, M. Hoesch, and R. J. Cava, arXiv:1507.04847.

[33] Z. Wang, M. G. Vergniory, S. Kushwaha, M. Hirschberger, E. V. Chulkov, A. Ernst, N. P. Ong, R. J. Cava, and B. A. Bernevig, Phys. Rev. Lett. **117**, 236401 (2016).

[34] T. Suzuki, R. Chisnell, A. Devalakonda, Y.-T. Liu, W. Feng, D. Xiao, J. W. Lynn, J. G. Chekelsky, Nat. Phys. **12**, 1119 (2016).

[35] A. Artmann, A. Mewis, M. Roepke, and G. Michels, Z. Anorg. Allg. Chem. **622**, 679 (1996).

[36] J.-R. Soh, C. Donnerer, K. M. Hughes, E. Schierle, E. Weschke, D. Prabhakaran, A. T. boothroyd, Phys. Rev. B **98**, 064419 (2018).

[37] A. F. May, M. A. McGuire, and B. C. Sales, Phys. Rev. B **90**, 075109 (2014).

[38] D. Shoenberg, *Magnetic Oscillations in Metals* (Cambridge University Press, Cambridge, UK, 2009.)

[39] E. M. Lifshits, A. M. Kosevich, J. Phys. Chem. Solids **4**, 1 (1958).

[40] L. P. He, L. P. He, X. C. Hong, J. K. Dong, J. Pan, Z. Zhang, J. Zhang, and S. Y.

Li, Phys. Rev. Lett. **113**, 246402 (2014)

[41] J. Park, G. Lee, F. Wolff-Fabris, Y. Y. Koh, M. J. Eom, Y. K. Kim, M. A. Farhan, Y. J. Jo, C. Kim, J. H. Shim, J. S. Kim, Phys. Rev. Lett. **107**, 126402 (2011).

[42] G. Mikitik, Y. V. sharlai, Phys. Rev. Lett. **82**, 2147 (1999).

[43] I. A. luk'yanchuk, Y. Kopelevich, Phys. Rev. Lett. **93**, 166402 (2004).

[44] I. A. luk'yanchuk, Y. Kopelevich, Phys. Rev. Lett. **97**, 256801 (2006).

[45] A. A. Taskin, Y. Ando, Phys. Rev. B **84**, 035301 (2011).

[46] H.-Z. Lu, S.-Q. Shen, Front. Phys. **12**, 127201 (2017).

[47] G. Hua, S. Nie, Z. Song, R. Yu, G. Xu, and K. Yao, Phys. Rev. B 98, 201116(R) (2018).

[48] M. C. Rahn, J.-R. Soh, S. Francoual, L. S. I. Veiga, J. Strempfer, J. Mardegan, D. Y. Yan, Y. F. Guo, Y. G. Shi, and A. T. Boothroyd, Phys. Rev. B **97**, 214422 (2018).

[49] H. J. Kim, K. S. Kim, J. F. Wang, M. Sasaki, N. Satoh, A. Ohnishi, M. Kitaura, M. Yang, and L. Li, Phys. Rev. Lett. **111**, 246603 (2013).

[50] A. Sakai, *et al*. Nat. Phys. **14**, 1119 (2018).

[51] Y. Ferreiros, A. A. Zyuzin, and J. H. Bardarson, Phys. Rev. B **96**, 115202 (2017).

[52] O. V. Kotov, and Y. E. Lozovik, Phys. Rev. B **98**, 195446 (2018).

[53] J. H. Wilson, A. A. Allocca, and V. Galitski, Phys. Rev. B **91**, 235115 (2015).

[54] P. E. C. Ashby, and J. P. Carbotte, Phys. Rev. B **89**, 245121 (2014).

[55] S. A. Parameswaran, T. Grover, D. A. Abanin, D. A. Pesin, and A. Vishwanath, Phys. Rev. X **4**, 031035 (2014).

[56] L.-L. Wang, N. H, Jo, B. Kuthanazhi, Y. Wu, R. J. McQueeney, A. Kaminski, and P. C. Canfield, arXiv: 1901.08234.

[57] J.-R. Soh, F. de Juan, M. G. Vergniory, N. B. M. Schröter, M. C. Rahn, D. Y. Yan, Y. F. Guo, Y. G. Shi, T. K. Kim, S. H. Simon, Y. Chen, A. I. Coldea, A. T. Boothroyd, arXiv: 1901.10022.

[58] P. E. Blöchl, Phys. Rev. B **50**, 17953 (1994); G. Kresse and D. Joubert, Phys. Rev. B **59**, 1758 (1999).

[59] G. Kresse and J. Hafner, Phys. Rev. B **47**, 558 (1993); G. Kresse and J. Furthmüller, Comp. Mater. Sci. **6**, 15 (1996); Phys. Rev. B **54**, 11169 (1996).

[60] J. P. Perdew, K. Burke, and M. Ernzerhof, Phys. Rev. Lett. **77**, 3865 (1996).

[61] S. L. Dudarev, G. A. Botton, S. Y. Savrasov, C. J. Humphreys, and A. P. Sutton, Phys. Rev. B **57**, 1505 (1998).

[62] N. Marzari and D. Vanderbilt, Phys. Rev. B **56**, 12847 (1997).

[63] I. Souza, N. Marzari, and D. Vanderbilt, Phys. Rev. B **65**, 035109 (2001).

**Acknowledgements**

The authors acknowledge the support by the Natural Science Foundation of Shanghai (Grant No. 17ZR1443300), the ShanghaiTech University startup fund, the Shanghai Pujiang Program (Grant No. 17PJ1406200), the National Key R&D Program of China (Grant No. 2017YFA0302903, 2017YFA0305400) and the National Natural Science Foundation of China (Grant No. 11774424, 11674229). The high magnetic field experiment was carried out at Wuhan National High Magnetic Field Center. Gong and Liu wish to thank Zhong-Yi Lu and Jian-Feng Zhang for helpful discussions. Computational resources were supported by Computational resources were provided by the Physical Laboratory of High Performance Computing at Renmin University of China, the HPC Platform of ShanghaiTech University Library and Information Services, as well as School of Physical Science and Technology.

**Author contributions**

Y.F.G. conceived and planned the experimental project. H.S. H.Y.W. and W.X. grew the crystals and prepared the samples. H.S., Z.H.Y. and N.Y. contributed to single crystal x-ray diffraction chracterizations on the structure and quality of the crystals. H.S., X.W. and Z.Q.Z. performed magnetization and electrical resistivity measurements. H.S. Z.W.Z., J.H.W., L.C.D., L.C.X., and X.K.L. conducted high magnetic field magneto-transport measurements and analyzed the data. H.F.Y., Z.K.L. and Y.L.C. provided the ARPES data. The *ab initio* calculations were carried out by B.C.G., W.J.S., P.G.G., K.L., and G.L. G.L. mapped out the Fermi arcs based on calculations. H.S., Y.F.G., G.L. and K.L. wrote the paper with inputs from all coauthors. All authors discussed the results and commented on the manuscript.

**Competing financial interests**

The authors declare no competing financial interests.

**Correspondence**

Correspondence and requests for materials should be addressed to Z.K.L., K.L., G.L. and Y.F.G..

**Figure 1 High-field magneto-transport data of EuCd$_2$Sb$_2$.** (a) The longitudinal resistivity $\rho_{xx}$ measured up to B = 55 T with B//$c$⊥I between T = 2 – 30 K. The inset shows the transverse Hall resistivity $\rho_{xy}$ versus magnetic field measured at 2 K. (b) The SdH oscillatory component as a function of 1/B obtained after subtracting a smooth background. (c) The corresponding FFT spectra of the quantum oscillations. Two fundamental frequencies of 30 T and 117 T can be identified. (d) The temperature dependence of relative amplitudes of the SdH oscillations. The solid line denotes the L-K formula fitting. The inset shows the Landau-level indices extracted from the SdH oscillations plotted as a function of 1/B. The solid line indicated a linear plot to the data.

**Figure 2 Electronic Structures of EuCd$_2$Sb$_2$ in the AFM and FM state and the Weyl points in the Brillouin zone.** (a) Calculated band structure of EuCd$_2$Sb$_2$ in the AFM state along the high-symmetry paths of Brillouin zone with the Hubbard interaction U among Eu 4$f$ electrons (U = 4.5 eV) and the spin-orbit coupling (SOC) included. (b) Details of the hole bands inversion around the Fermi level. (c) The tiny gaps along Γ-Z direction opened by the $C_3$ symmetry broken caused by the AFM structure. (d) Calculated band structure of EuCd$_2$Sb$_2$ in the FM state with similar conditions as those used for the AFM state. (e) The five pairs of Weyl points emerged along Γ-Z direction of BZ within ±0.1 eV. The black and blue colors denote opposite chirality of these WPs.

**Figure 3 Low-field resistivity and chiral anomaly of EuCd$_2$Sb$_2$.** (a) The resistivity $\rho_{xx}$ (black solid line) and $\rho_{zz}$ (red solid line) at low magnetic fields with B//$c$⊥I and B//$c$//I at 2 K, respectively. The inset shows an enlarged view of the data between 4 T and 8 T. (b) Results of the fitting to $\rho_{zz}$ at 2 K between -0.6 T – 0.6 T by the chiral anomaly equation.

**Figure 4 Fermi arcs on the (100) and (010) surfaces of EuCd$_2$Sb$_2$.** The surface states of EuCd$_2$Sb$_2$ at constant energy with values of (a, b) E = Ef and (c, d) E = Ef-0.06 eV. The two

columns, i.e. (a, c) and (b, d), correspond to the states at (100) and (010) surfaces with the surface BZ indicated by light-red and light-blue surfaces in the middle subset. Inside each plot, the locations of the projected bulk WPs are marked with the same color code as used in Figure 3 (e) to represent their chirality. Mirrored by the horizontal axis, these WPs form pairs with possible Fermi arc connected in between.

**Figure 1**

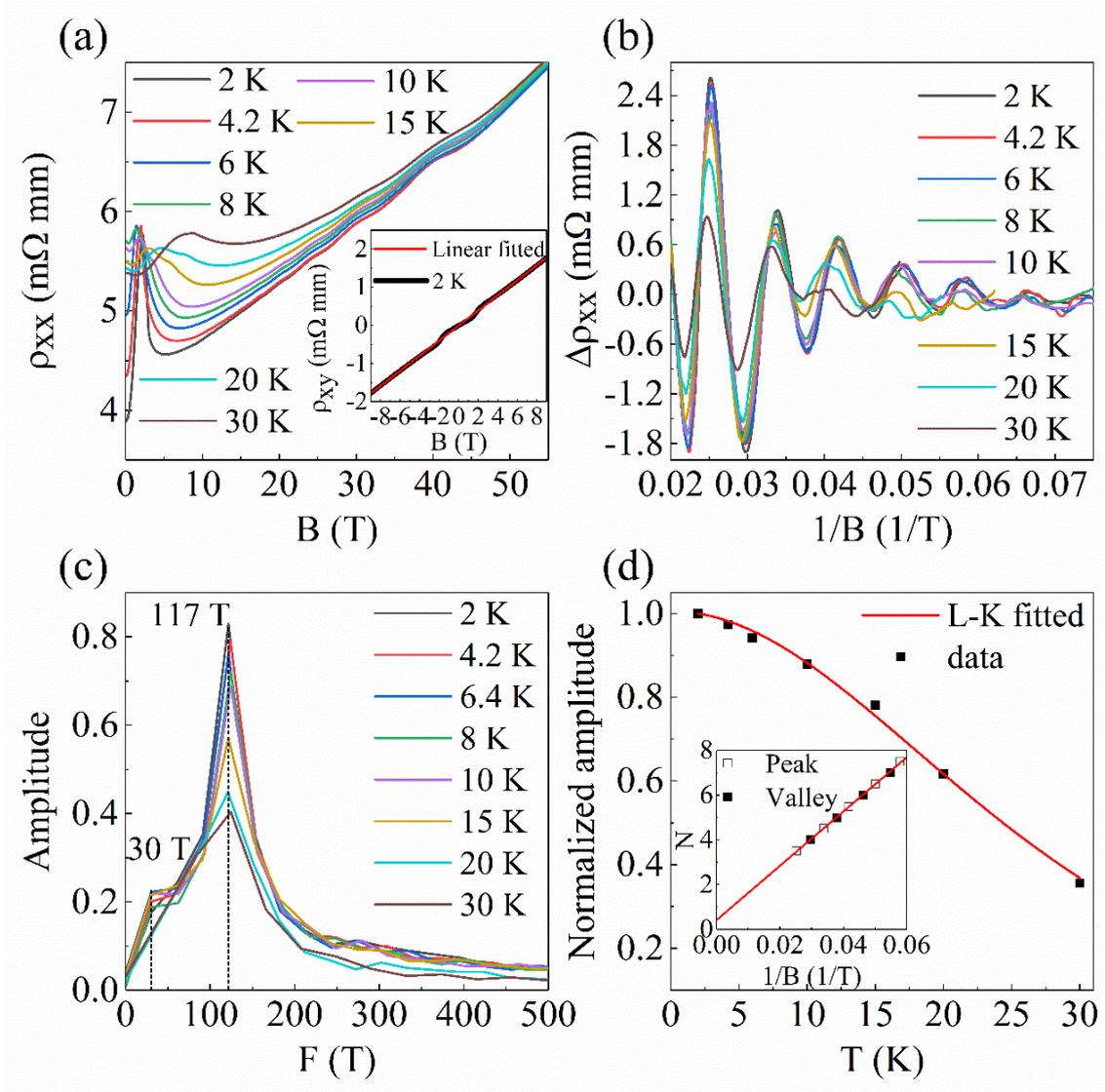

**Figure 2**

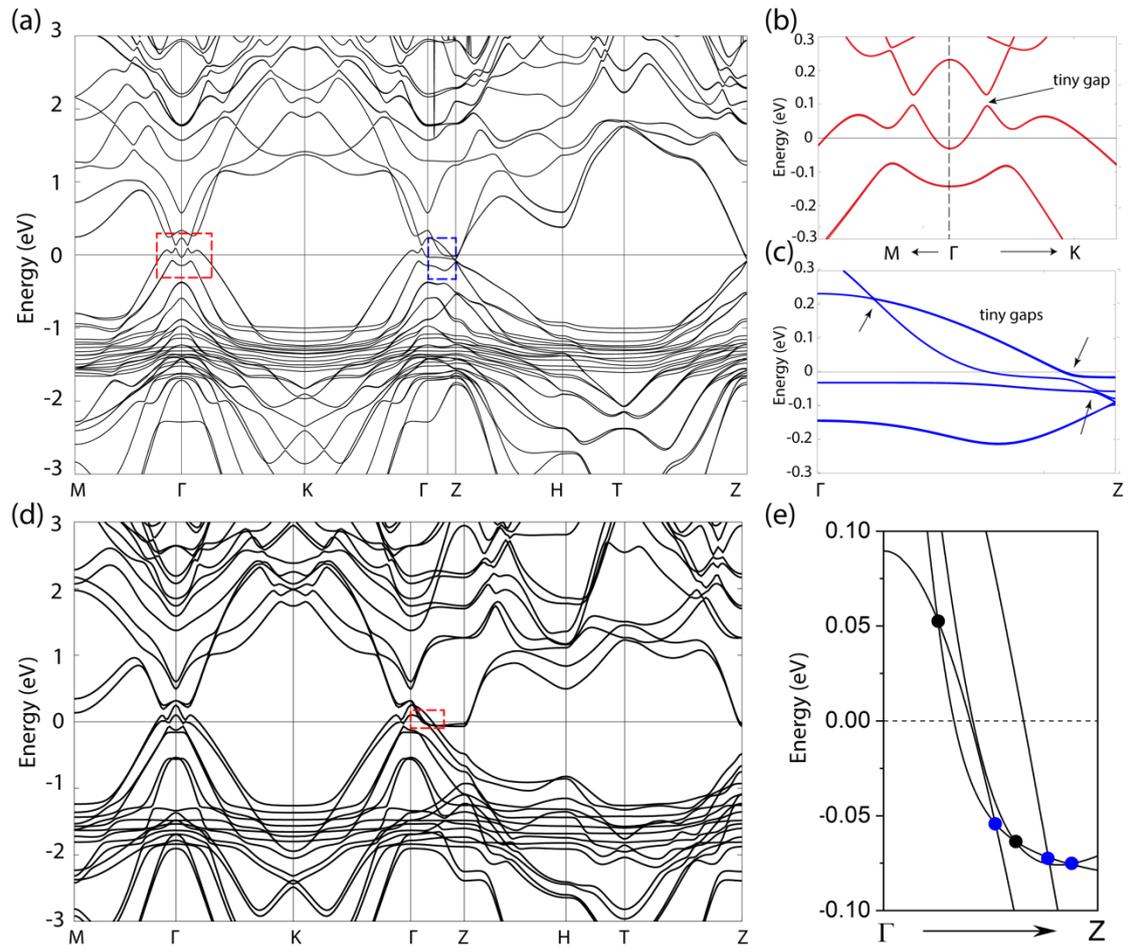

**Figure 3**

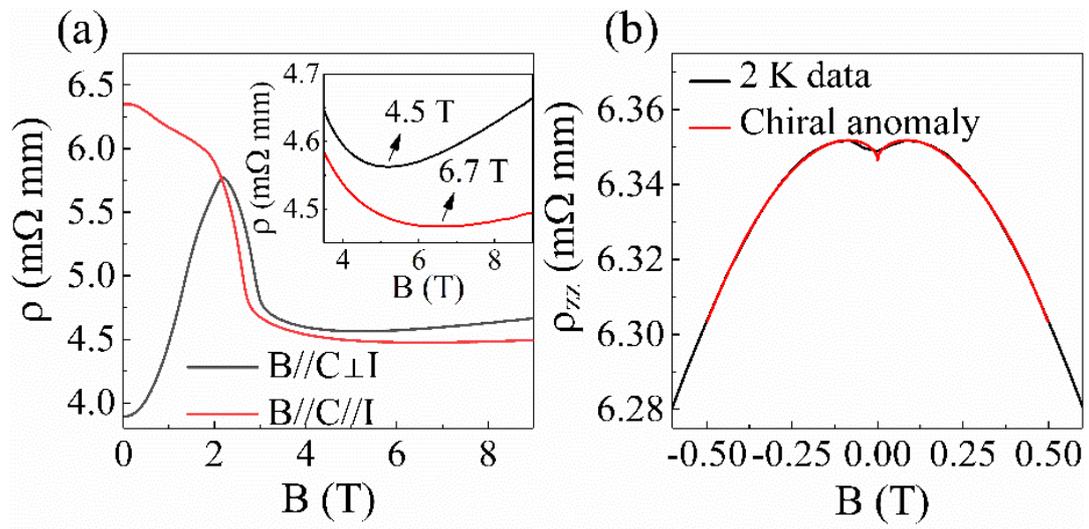

**Figure 4**

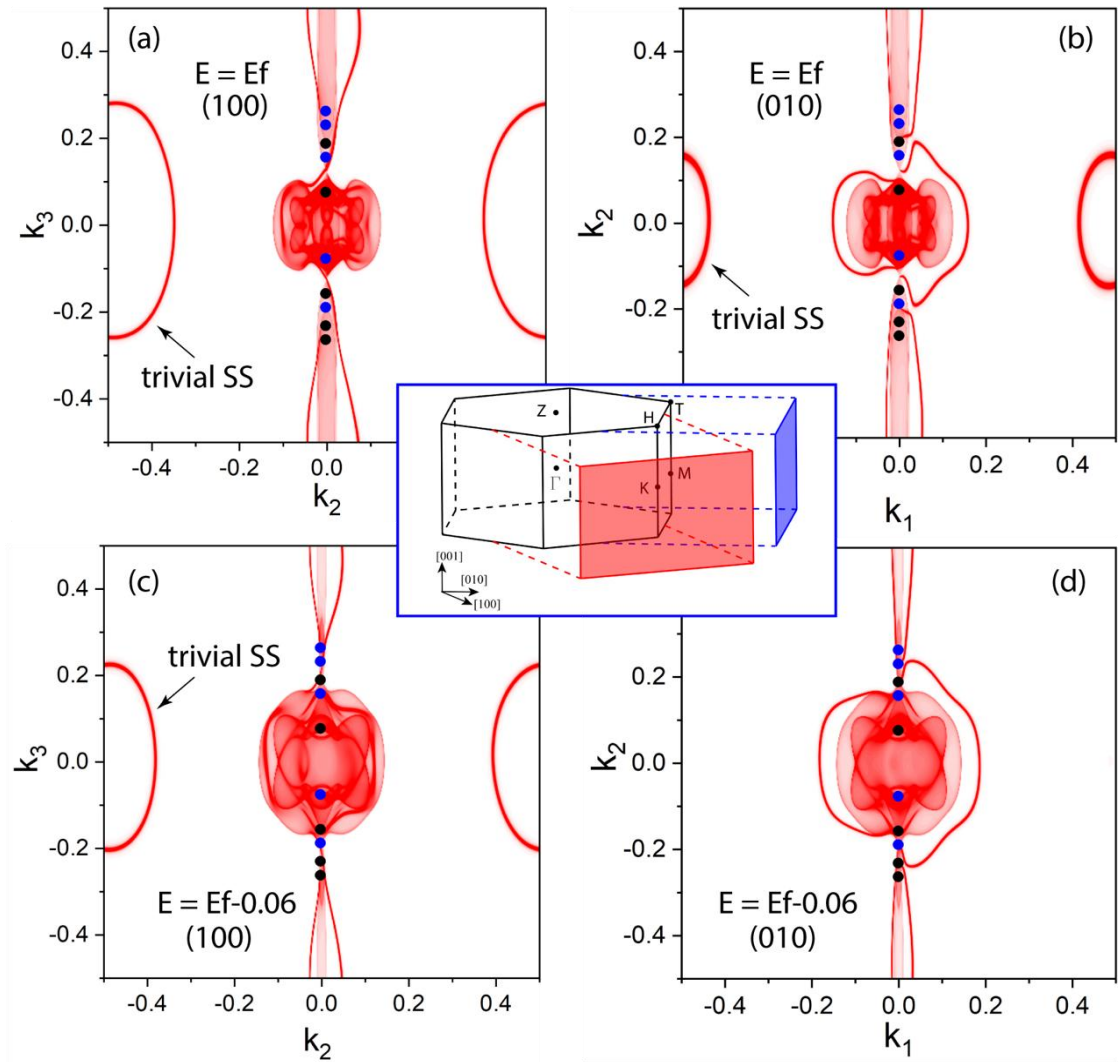

# Supplementary Information

# Magnetic exchange induced Weyl states in a semimetal EuCd$_2$Sb$_2$

Hao Su[1,2,†], Benchao Gong[3,†], Wujun Shi[1,†], Haifeng Yang[1], Hongyuan Wang[1,2], Zhenhai Yu[1], Peng-Jie Guo[3], Wei Xia[1,2], Jinhua Wang[4], Linchao Ding[4], Liangcai Xu[4], Xiaokang Li[4], Xia Wang[5], Zhiqiang Zou[5], Na Yu[5], Zengwei Zhu[4], Yulin Chen[1,6], Zhongkai Liu[1*], Kai Liu[3*], Gang Li[1*], Yanfeng Guo[1,2*]

[1]School of Physical Science and Technology, ShanghaiTech University and CAS-Shanghai Science Research Center, Shanghai 201210, China
[2]University of Chinese Academy of Sciences, Beijing 100049, China
[3]Department of Physics and Beijing Key Laboratory of Opto-electronic Functional Materials & Micro-nano Devices, Renmin University of China, Beijing 100190, China
[4]Wuhan National High Magnetic Field Center and School of Physics, Huazhong University of Science and Technology, Wuhan 430074, China
[5]Analytical Instrumentation Center, School of Physical Science and Technology, ShanghaiTech University, Shanghai 201210, China
[6]Department of Physics, University of Oxford, Oxford, OX1 3PU, UK

[†]The authors contributed equally to this work.
[*]E-mails:
liuzhk@shanghaitech.edu.cn;
kliu@ruc.edu.cn;
ligang@shanghaitech.edu.cn;
guoyf@shanghaitech.edu.cn.

### a. Crystal chracterizations

Clean single crystals with typical size of 1.2 ×1 ×0.2 mm were obtained, shown by a representative picture in Figure S1(a). Compositions of the $EuCd_2Sb_2$, as are demonstrated by the EDX, are rather close to the nominal values and the Sn flux was not detected. The prefect reciprocal space lattice without any other miscellaneous points, seen in Figures S1(e)-S1(g), indicates pure phase and high quality of the crystal used in this study. The diffraction pattern could be satisfyingly indexed on the basis of a trigonal structure with lattice parameters $a$ = 4.6926(6) Å and $c$ =7.7072(8) Å in the space group *P*-3*m*1 (No. 164), consistent with those values reported previously[1]. Based on a careful refinement, the crystal structure was precisely solved as shown in Figure S1(b) (the arrows denote the spin directions) in which two Sb-Cd zigzag chains are interlaced and superimposed along the *c*-axis between two Eu atom layers. The views of structure from other different directions are also illustrated in Figures S1(c) and S1(d).

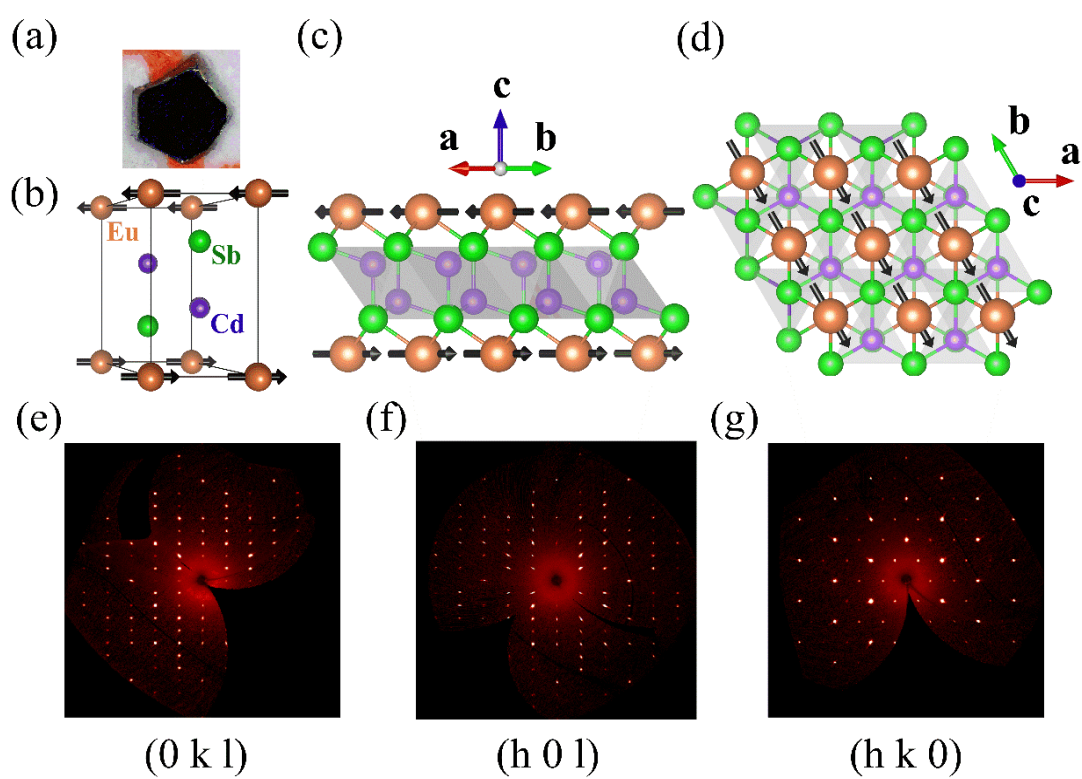

**Figure S1**. (a) The SEM image for a typical $EuCd_2Sb_2$ crystal. (b)-(d) The sketch views of crystal structures of EuCd$_2$Sb$_2$ along *a*, *b*, and *c* orientations, respectively. (e)-(g) The diffraction patterns in the reciprocal space along (0kl), (h0l), and (hk0) directions, respectively.

b. **Magnetization and resistivity measurements.**

The temperature dependent magnetization (M(T)) along the out-of-plane (B//*c*) direction is shown in Figure S2(a). The Curie-Weiss plot to the paramagnetic region of M(T) above 50 K gave the Weiss temperature of $\theta_p$ = -4.6(5) K and an effective magnetic moment of about 7.94 $\mu_B$ which is consistent with the expectation from Eu$^{2+}$(4$f^7$, $S = 7/2$, $L = 0$). These values are also well consistent with those previously reported[1]. An anomaly is observed at $T_N \sim 7.5$ K, signifying the AFM ordering at this temperature. By increasing the external magnetic field, the AFM order is apparently suppressed, consistent with the fact that the spins of Eu$^{2+}$ are fully polarized along the *c*-axis and the system eventually enters into the FM state when B > 3.2T. The inset in Figure S2(a) shows saturation moment at 2 K close to 7 $\mu_B$, suggesting that spins of the localized Eu 4*f* electrons are actually saturated. It is obviously that the saturation field with respect to the out-of-plane direction is about 3.2 T.

The temperature dependence of longitudinal resistivity $\rho_{xx}$ at various B presented in Figure S2(b) shows several prominent features: above 40 K $\rho_{xx}$ displays an almost linear temperature dependence with the residual resistivity ratio $\rho_{xx}$(300K)/$\rho_{xx}$(40K) being approximately 1.5, indicative of an essential semi-metallic behavior. Below 40 K, $\rho_{xx}$ starts to gradually increase upon cooling till it reaches a peak at $T_N$, and subsequently exhibits a sudden drop with further decreasing the temperature to 2 K. The low temperature behavior of $\rho_{xx}$, totally resembling that of the sister compound EuCd$_2$As$_2$, could be reasonably understood in connection with the scattering of conduction electrons by the localized Eu$^{2+}$ which is coupling with the Cd and As orbitals and the gradual spin alignment in external magnetic field. The disappearance of such peak at B > 3.2T which is the critical magnetic field (B$_c$) at which the spins are completely aligned along *c* axis, also supports this argument.

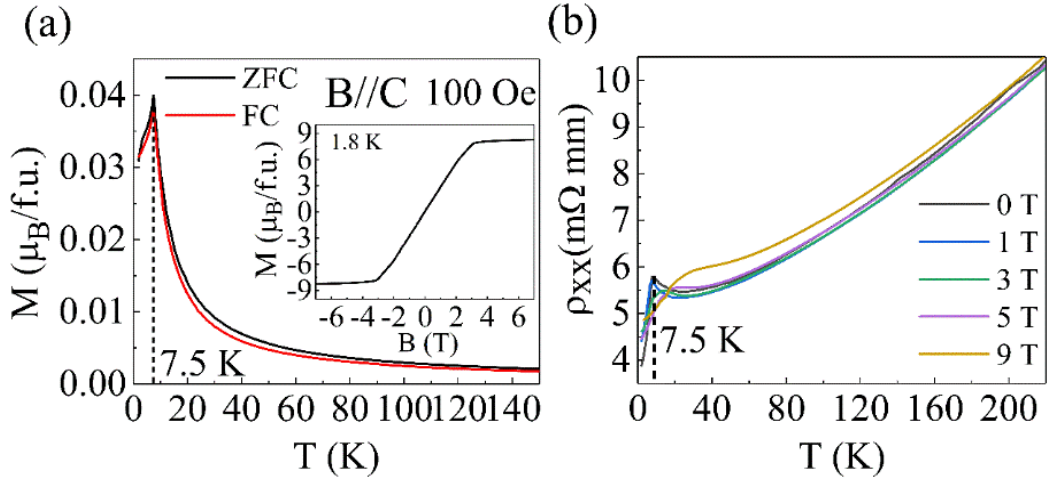

**Figure S2**. (a) The temperature dependence of magnetizations of $EuCd_2Sb_2$ measured with B = 100 Oe and the inset is the isothermal magnetization measured at 1.8 K between -7 T – 7 T. (b) The temperature dependence of longitudinal resistivity $\rho_{xx}$ measured at various magnetic fields up to 9 T.

## c. High magnetic field magneto-transport measurements at I//B//c configuration

Figure S4 presents the longitudinal MR $\rho_{zz}$ measured with B//c//I and the analysis yielded results consistent with those derived from $\rho_{zz}$, suggesting the reliability of our analysis.

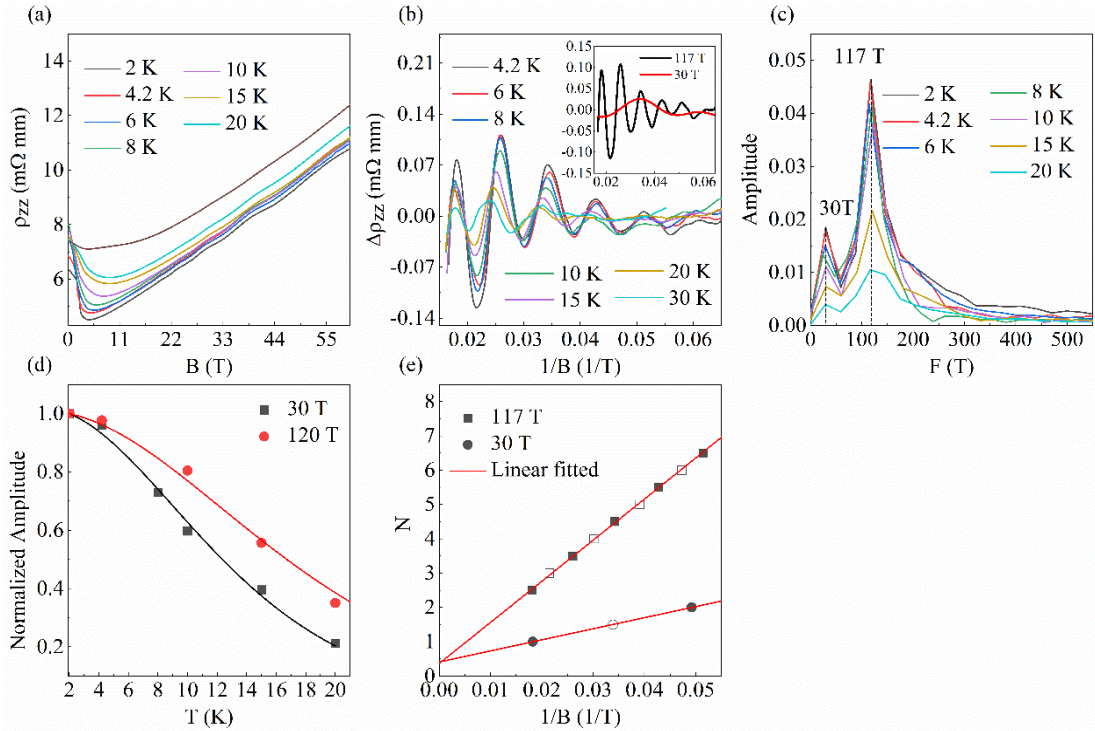

**Figure S4.** (a) The resistivity $\rho_{zz}$ measured up to B = 60 T with B//$c$//I between T = 2 – 20 K. (b) The SdH oscillatory component as a function of 1/B obtained after subtracting a smooth background. The inset shows the data for the two frequencies of 30 T and 117 T. (c) the corresponding FFT spectra of the quantum oscillations, which also reveal two fundamental frequencies of 30 T and 117 T. (d) The temperature dependence of relative amplitudes of the SdH oscillations. The solid lines denote the L-K formula fitting. (e) The Laudau-level indices extracted from the SdH oscillations plotted as a function of 1/B. The solid line indicated the linear plots to the data.

### d. ARPES measurements and the Hubbard U determination

ARPES data was mainly acquired at BL4 and BL10 of ALS, USA, at 20 K (above the magnetic transition temperature of EuCd$_2$Sb$_2$, where the strong short-range FM correlations persist). The overall energy and angle resolutions were 15 meV and 0.2°, respectively. Fresh EuCd$_2$Sb$_2$ surfaces for ARPES measurements were obtained by *in-situ* cleaving the crystals at low temperatures (20 K). Figure S5 plots the measured band dispersion along the high symmetry $\overline{K}$-$\overline{\Gamma}$-$\overline{K}$ direction, which contains the dispersive bands near $E_F$ and flat Eu 4f bands sitting at 1.7 eV below $E_F$. The FS is dominated by hole type carriers and the predicted WPs are consequently difficult to be observed. By comparing with the *ab initio* calculations with different Hubbard interaction U in the FM phase, we found the best agreement at U = 4.5 eV. Despite the absence of long-range magnetic order, the persisting short-range FM correlations render the electronic structure measured in ARPES a great similarity to the one calculated in the FM phase. Therefore, ARPES measurements provide experimental proof for the choice of a reasonable U.

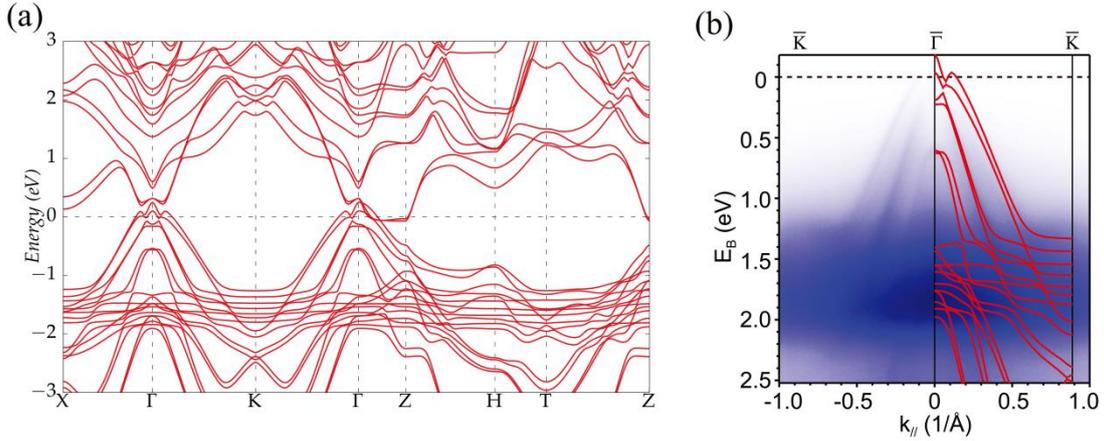

**Figure S5**. (a) Calculated band structure of EuCd$_2$Sb$_2$ in **FM** state along the high-symmetry paths of Brillouin zone (BZ) with the Hubbard interaction U among Eu 4f electrons (U = 4.5 eV) and the spin-orbit coupling (SOC). (b) The APRES result along the $\bar{K}$-$\bar{\Gamma}$-$\bar{K}$ path superimposed with the calculated band structure.

e. *ab initio* calculations

The first-principles electronic structure calculations on EuCd$_2$Sb$_2$ were carried out by using the projector augmented wave (PAW) method[3] as implemented in the VASP package[4]. The generalized gradient approximation (GGA) of Perdew-Burke-Ernzerhof[5] was employed for the exchange-correlation functional. The kinetic energy cutoff of the plane-wave basis was set to be 340 eV. The experimental lattice constants and atomic positions of EuCd$_2$Sb$_2$ were adopted[1]. A $16 \times 16 \times 10$

*k*-point mesh for the Brillouin zone (BZ) sampling of the primitive cell and the Gaussian smearing technique with a width of 0.05 eV for the FS broadening were utilized. The spin-orbit coupling (SOC) effect was taken into consideration. The correlation effect among Eu 4*f* electrons was incorporated by using the GGA+U formalism of Dudarev *et al.*[6] with an effective Hubbard interaction U. According to a previous study[1], the energies of Eu 4*f* bands relative to the $E_F$ are sensitive to the Hubbard interaction. We herein chose the Hubbard U of 4.5 eV as is discussed in the ARPES measurements. The maximally localized Wannier functions method[7,8] was used to calculate the FSs, from whose volumes the carrier concentrations were obtained.

To further validate the choice of the Hubbard interaction parameters, we further studied the magnetic properties of $EuCd_2Sb_2$ in several typical magnetic configurations of Eu spins as in Figure S6. Depending on the Hubbard interaction strength between Eu *f*-electrons, three different types of collinear magnetic states can be stabilized. The A-type AFM configuration was found to be the most energetic favorable state for U between 3 and 8 eV, which validates the choice of effective U value (4.5 eV) chosen from the comparison to ARPES measurement (Figure S5). The magnetic ground state of $EuCd_2Sb_2$ in A-AFM configuration with weak coupling of inter-layer Eu spins also agrees with the experimentally measured Neel temperature of 7.4 K[1] and is similar to the case of $EuCd_2As_2$[3,4]. Further calculations with different easy axis of magnetization show that Eu spins prefer to lie in the *ab*-plane. Nevertheless, the magnetic anisotropy energy (MAE), i.e., the energy difference (ΔE) between the in-plane and out-of-plane spin directions, is very small (ΔE = 0.2 meV/Eu). The weak inter-layer interaction as well as the small MAE indicate that $EuCd_2Sb_2$ is prone to the external magnetic field, which is in good accordance with above low saturation field in the magnetization measurements, seen by the inset in Figure S2(a).

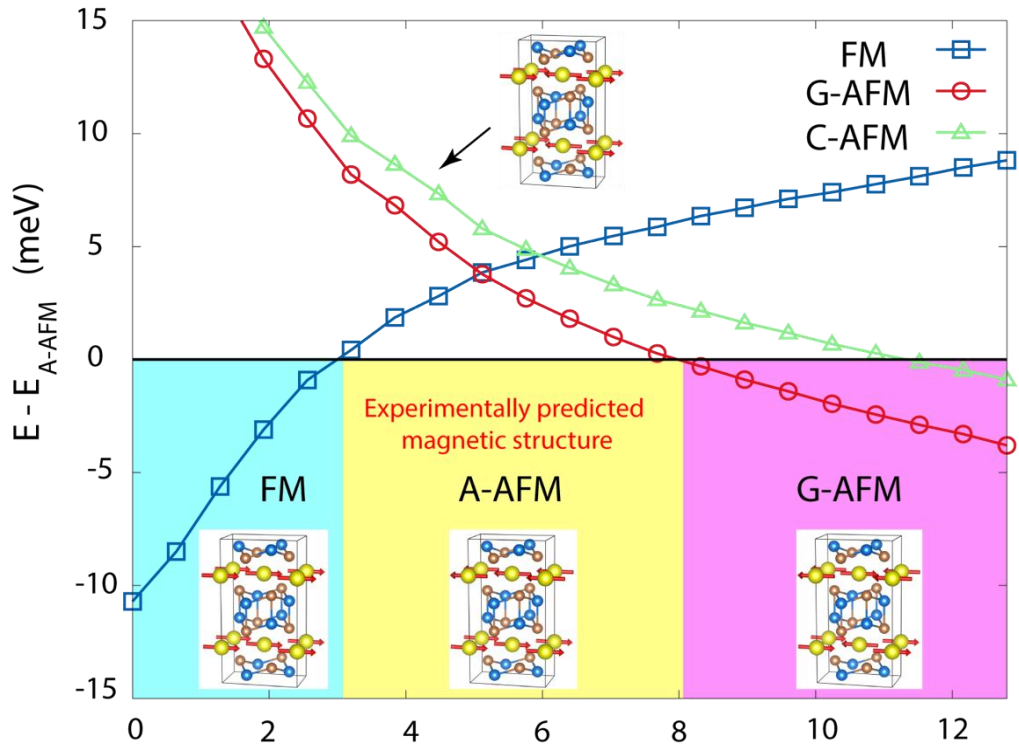

**Figure S6**. Magnetic phase diagram of $EuCd_2Sb_2$ as a function of Hubbard U. The collinear type FM, A-AFM, C-AFM and G-AFM configurations have been examined. The A-AFM is found to be more stable than other magnetic configurations over a large range of Hubbard interactions, i.e. (-3, 8) eV. As shown in Fig. S3, U = 4.5 eV is experimentally favored which nicely resides in the A-AFM phase in agreement with our magnetic measurements.

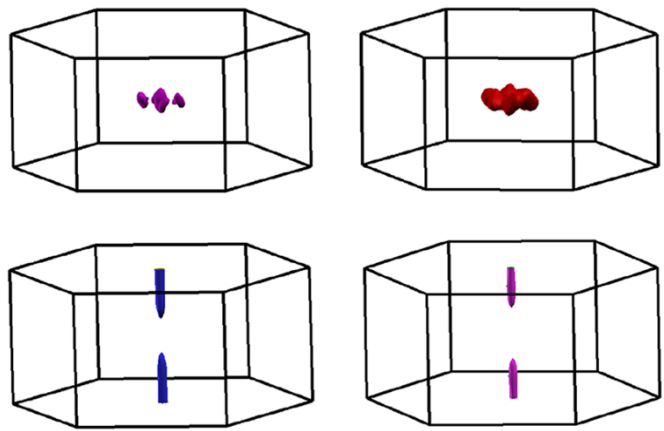

**Figure S3**. The FSs of $EuCd_2Sb_2$ are contributed by four bands with only two geometrically distinct shapes in the FM state. The FSs were prepared with the XCRYSDEN program[2].

Table SI. The parities of all filled states of AFM $EuCd_2Sb_2$ at eight time-reversal invariant momentum points in the Brillouin zone.

|  | Γ | M | M | M | A | L | L | L | Total |
|---|---|---|---|---|---|---|---|---|---|
| parities | + | − | − | − | + | + | + | + | − |

Table S2. The positions and energies of five pairs of Weyl points in the BZ. The $(k_1,k_2,k_3)$ are positions in reciprocal lattice of unit cell.

| No. | $k_1$ | $k_2$ | $k_3$ | $E - E_f\ (meV)$ |
|---|---|---|---|---|
| $W_1^\pm$ | 0 | 0 | ±0.075 | +49.49 |
| $W_2^\pm$ | 0 | 0 | ±0.160 | −62.48 |
| $W_3^\pm$ | 0 | 0 | ±0.1984 | −71.85 |
| $W_4^\pm$ | 0 | 0 | ±0.2322 | −77.19 |
| $W_5^\pm$ | 0 | 0 | ±0.2422 | −78.47 |

**References**

[1] J.-R. Soh, C. Donnerer, K. M. Hughes, E. Schierle, E. Weschke, D. Prabhakaran, A. T. boothroyd, Phys. Rev. B **98**, 064419 (2018).

[2] A. Kokalj, Comp. Mater. Sci. **28**, 155 (2003).

[3] G. Hua, S. Nie, Z. Song, R. Yu, G. Xu, and K. Yao, Phys. Rev. B 98, 201116(R) (2018).

[4] M. C. Rahn, J.-R. Soh, S. Francoual, L. S. I. Veiga, J. Strempfer, J. Mardegan, D. Y. Yan, Y. F. Guo, Y. G. Shi, and A. T. Boothroyd, Phys. Rev. B **97**, 214422 (2018).